\documentstyle[aps,preprint]{revtex}

\newcommand{\bea}{\begin{eqnarray}}
\newcommand{\eea}{\end{eqnarray}}
\newcommand{\bear}{\begin{eqnarray*}}
\newcommand{\eear}{\end{eqnarray*}}

\begin{document}
\draft 
\title{
NEW INTEGRABLE MODELS OF STRONGLY CORRELATED PARTICLES WITH CORRELATED HOPPING}
\author{
F. C. Alcaraz$^1$ and R. Z. Bariev$^{1,2}$}
 
\address{$^1$Departamento de F\'{\i}sica,
Universidade Federal de S\~ao Carlos, 13565-905, S\~ao Carlos, SP
Brazil}
 
\address{$^2$The Kazan Physico-Technical Institute of the Russian Academy of Sciences,
Kazan 420029, Russia}
 
\maketitle

\begin{abstract}
The exact solution is obtained for the eigenvalues and eigenvectors for two 
models of strongly correlated particles with single-particle correlated 
and uncorrelated pair hoppings. The asymptotic behavior of correlation functions 
are analysed in different regions, where the models exhibit different 
physical behavior.//

{Published in Phys. Rev. B,{\bf 59}, 3373 (1999)}

\end{abstract}
\pacs{PACS numbers: 75.10.Lp, 74.20-z, 05.50.+q }
%
\narrowtext

Integrable strongly correlated electron systems (see, e.g. [1]) have been an
important subject of research activity in recent years since they are 
believed to play a promising role in unraveling the mystery of high-$T_c$
superconductivity. Several integrable correlated fermion systems which
manifest superconducting properties so far appeared in the literature.
Most famous are the supersymmetric $t-J$ [2-4] model and the Hubbard model[5].
Other integrable correlated electron systems of interest include the
correlated hopping model [6] and its generalization 
extensively investigated in [7,8]. The models considered in these last papers
describe the dynamics of two type of particles (spin up and spin down) 
with kinetic terms given by correlated single particle hopping and 
uncorrelated hopping in the case of pair motion.

In this letter, we introduce two new integrable quantum chains  with correlated 
single-particle and uncorrelated pair hopping but having only 
one type of particles. 

 We suppose that particles on the
chain   may be isolated (both nearest neighbour sites are
empty) or  part of a two-atom molecule (one of its nearest neighbour sites is 
occupied and the other one is empty). This supposition is achieved by imposing 
the
restriction of no simultaneous occupancies of three nearest 
neighbor sites. It means 
that two particles on the neighbor sites create the two-atom molecule which 
can hop as a whole or disintegrate. The Hamiltonian of such system in the 
most general form can be presented as follows
\bea
H &=-& \sum_{j}{\cal P}\lbrace (\sigma_{j+1}^+\sigma_{j}^- + 
\sigma_j^+\sigma_{j+1}^-)\nonumber\\
&\times &[(1-n_{j-1})(1-n_{j+2}) + t_{1} n_{j-1}(1-n_{j+2}) \nonumber\\
&+& t_{2}
(1-n_{j-1})n_{j+2} 
+ t_{12}n_{j-1}n_{j+2}]\nonumber\\
&+& t_{p}(\sigma_{j+1}^+\sigma_{j-1}^- + \sigma_{j-1}^+\sigma_{j+1}^-)n_{j}
+ u n_{j}n_{j+1}\nonumber\\
 &+ &  n_{j-1}n_{j+1}[V_{11} + V_{12}n_{j+2} + V_{21}n_{j-2}
 \nonumber \\ &+& V_{22}n_{j-2}n_{j+2}]\rbrace {\cal P}
\eea
where $\sigma_j^+(\sigma_j^-) $ creates (annihilates) a particle at
site $j$ and 
\bea
n_{j} = \sigma_{j}^+\sigma_{j}^-\nonumber
\eea
is the corresponding occupation number.
The model contains correlated single-particle hopping, which is 
described by the parameters $t_1, t_2 , t_{12} $ and uncorrelated pair 
hopping described by the parameter $t_{p}$.
We have also two- , three- and four-particle static interactions between 
nearest-neighbors. The operator ${\cal P}$
 in (1) projects out   any configuration 
with simultaneous 
occupancies of three nearest neighbor sites and we assume periodic 
boundary conditions.

Certainly the Hamiltonian (1) is not integrable for an arbitrary 
choice of parameters. Therefore our problem is to find under what 
conditions the Hamiltonian (1) can be treated by the coordinate 
space Bethe ansatz technique.

The obvious way to identify a state of the Hamiltonian (1) with $n$
 particles is to specify their
positions $x_1,\ldots,x_n$ ordered so that 
\bea
1\le x_{1} \le x_{2}\le \ldots\le x_{n} \le N.
\eea

We assume the following ansatz for the wavefunction. If we have only 
particles $(x_{i+1}\ne x_{i}+1$, $i = 1,2,\ldots,n-1)$ we write
\bea
\Psi(x_{1},\ldots, x_{n}) = \sum_{P} A_{P_1\ldots P_n}
^{1\ldots1}\exp(i\sum_{j=1}^n{k_{P_j}x_{j}})
\eea
where $P$ is the 
permutation of $1,2,\ldots,n$ and ${k_P}$ are unknown quasiparticle momenta. 
The n superscript 1 in the amplitude indicates we have only isolated particles.
In the case we have a pair at the position $x_l,x_{l+1}$ $(x_{l+1}=x_l +1)$ the 
ansatz is 
\bea
\Psi(x_{1},\ldots, x_l,x_{l+1},\ldots, x_{n}) = \nonumber \\ 
\sum_{P} A_{P_1 \ldots 
P_{l}P_{l+1}\ldots P_n}
^{1\ldots\overline{11}\ldots 1}\exp(i\sum_{j=1}^n{k_{P_j}x_{j}})
\eea
where  the bar at the $l^{th}$ and $(l+1)^{th}$ position of the superscript 
indicates the pair's position. The general case with many isolated particles 
and pairs follows from eqs.(3) and (4).

It is not difficult  to consider the eigenvalue equations if $n=1,2$ or for
general $n$ in the case we have only isolated particles.
These equations give the connections between the coefficients 
$A_{P_1\ldots P_{n}}^{1\ldots 1}$.
\bea
 s_{P_j P_{j+1}}A_{\ldots P_jP_{j+1}\ldots }^{\ldots 11\ldots } + 
[P_j \leftrightarrow P_{j+1}]
 = 0 \nonumber\\
 N_{P_j P_{j+1}}^{(1)}A_{\ldots P_jP_{j+1}\ldots }^{\ldots 11\ldots } - 
 c_{P_j P_{j+1}}^{(1)}A_{\ldots P_jP_{j+1}\ldots }^{\ldots 11\ldots } + 
 [P_j \leftrightarrow P_{j+1}] = 0
\eea 
where
\bea
s_{P_1P_2} &=&N_{P_1 P_2}^{(1)}c_{P_1 P_2}^{(2)} - N_{P_1 P_2}^{(2)}
c_{P_1 P_2}^{(1)} \nonumber\\
N_{P_1 P_2}^{(1)}& =& 1 - e^{ik_{P_2}}V_{11} + e^{i(k_{P_1}+k_{P_2})}
 \nonumber\\
 N_{P_1 P_2}^{(2)} &=& t_2 e^{ik_{P_2}} + t_1e^{i(k_{P_1}+2k_{P_2})}
\nonumber\\
 c_{P_1 P_2}^{(1)} &=& t_1 + t_2e^{i(k_{P_1}+k_{P_2})}
\nonumber\\
 c_{P_1 P_2}^{(2)} &=& -t_p + e^{ik_{P_1}} + e^{ik_{P_2}} -
U e^{i(k_{P_1}+k_{P_2})}\nonumber\\
 &+& e^{i(2k_{P_1}+k_{P_2})}+e^{i(k_{P_1}+2k_{P_2})}
-t_pe^{2i(k_{P_1}+k_{P_2})}.
\eea
In contrast to the $XXZ$ model [2,9], this is not sufficient to prove 
that the  Bethe ansatz works. In order to do that we must consider the
eigenvalue equations
at the boundary of  the inequalities (2) for the case $n=3$ and 4. 
This  gives a complicated 
system of equations for the coupling constants of the Hamiltonian (1) .
We have treated this system
on a computer and found the following integrable cases 
\bea
(1) \hspace{1cm}t_1 &=& t_2 = 2\cosh\eta,\;\;  
t_{12} = 1, \;\; t_{p} =-\varepsilon,  \nonumber\\
 V_{11} &=&  
V_{22} = 0, U = 2t_{p} ;\; \; V_{12} = V_{21}  = \varepsilon(t_1^2 - 1),
\eea

and
\bea
(2) \hspace{1cm}t_{1} &=& \frac{\sinh 2\eta}{\cosh 3\eta}e^{2\eta}, \;\;
t_{2} = -\frac{\sinh 2\eta}{\cosh 3\eta}e^{-2\eta}, 
t_{12} = 1,
 t_{p}= \frac{\varepsilon\cosh\eta}{\cosh 3\eta}; \nonumber\\
 V_{11} &=& 2t_{p}\cosh2\eta ; \;\;
U = 2t_p(1 + 4\sinh^2\eta\cosh 2\eta); \nonumber\\
V_{12} &=& (e^{-4\eta} - 2\cosh2\eta)t_{p}; \;\;V_{21} = 
(e^{4\eta} - 2\cosh2\eta)t_{p}; \nonumber \\ V_{22}& =& 0,
\eea
where $\varepsilon = \pm 1$.

The Bethe-ansatz equations are derived following the standard procedure [1-2].
Each state of the Hamiltonian is specified by a set of particle rapidities
$\lambda_j(j=1,\ldots ,n) $related to the momenta of particles $ k_j$.
The rapidities have to satisfy the Bethe-ansatz equations. For the model (7)
these equations have the following form
\bea
\left[\frac{\varepsilon\sin(\lambda_j +i\eta)}
{\sin(\lambda_j -i\eta)}\right]^{N-n} &=& e^{-iP}\prod_{l=1}^n
\frac{\sin(\lambda_j -\lambda_l +i\eta)}
{\sin(\lambda_j -\lambda_l-i\eta)}\\
e^{ik_j} &=& \varepsilon \frac{\sin(\lambda_j +i\eta)} {\sin(\lambda_j -i\eta)}
\eea

and for the model (8) these equations are 
\bea
\left[\frac{\varepsilon\sin(\lambda_j -i\eta)}
{\sin(\lambda_j +i\eta)}\right]^{N-n} &=& e^{-iP}\prod_{l=1}^n
\frac{\cos(\lambda_j -\lambda_l +i\eta)}
{\cos(\lambda_j -\lambda_l-i\eta)}
\frac{\sin(\lambda_j -\lambda_l -2i\eta)}
{\sin(\lambda_j -\lambda_l+2i\eta)}\\
e^{ik_j} &=& \varepsilon \frac{\sin(\lambda_j -i\eta)}
{\sin(\lambda_j +i\eta)}
\eea

In both cases 

\bea
P=\sum_{l=1}^n k_l
\eea
is the momentum and the energy of the system is given by

\bea
E=-2\varepsilon\sum_{j=1}^n\left[\cosh2\eta - \frac{\sinh^{2}2\eta}
{\cosh2\eta-\cos2\lambda_j}\right].
\eea
It is interesting to observe the similarity of the Bethe ansatz equations 
(9) and (11), and those of the spin-1 Zamolodchikov-Fateev model [10] and
Izergin-Korepin model[11], respectively.
Although both  models are exactly integrable , let us restrict ourselves to 
the more physically interesting model (7) with density $\rho < 1/2$. 
There are different regions, with distinct physical properties depending 
on the parameter $\Delta = -\varepsilon\cosh\eta $.

1) $\Delta = < -1, \varepsilon = +1, \eta$ is real .
We have a similar situation as in the ferromagnetic non-critical $XXZ$-chain
with fixed magnetization [2,12]. The ground state must contain 
exactly one string of maximum length. It means that there is a gap for the
arbitrary concentration of particles and the system is in a phase separated 
region, where all the particles prefer to stay together.

2) $\Delta > 1, \varepsilon = -1, \eta$ is real.
We may analyse this case by considering the limiting case $\eta 
\rightarrow + \infty$. From this analysis 
it is clear that the 
ground state contains $n/2$ bound pairs characterized by 
a pair of complex particle rapidities
\bea
\lambda_{\alpha}^{\pm} = \frac{1}{2}(\lambda_{\alpha}^{(2)} \pm i\eta).
\eea
Inserting (15) in eqs.(10) and introducing the density function
$\rho(\lambda)$ for the distribution of $\lambda_{\alpha}^{(2)}$
in the thermodynamic limit, we obtain the linear integral equation
\bea
2\pi\rho(\lambda) +\int_{-\lambda_0}^{\lambda_0}
{[2\Phi^{(1)}(\lambda - \lambda';2\eta) + \Phi^
{(1)}(\lambda-\lambda';4\eta)]
\rho(\lambda')}d\lambda'= \nonumber\\
(1 -\frac{n}{N})[\Phi^{(1)}(\lambda,\eta) + \Phi^{(1)}(\lambda, 3\eta)],
\eea
where
\bea
\Phi^{(1)}(\lambda,\eta) = \frac{\sinh\eta}{\cosh\eta -\cos\lambda}.
\eea
The parameter $\lambda_0$ is determined by the subsidiary condition for 
the total density $\rho = n/N$ of particles
\bea
\int_{-\lambda_0}^{\lambda_0}\rho(\lambda)d\lambda = \frac{1}{2}\rho.
\eea
Similar calculations gives us the ground-state energy 
\bea
\frac{1}{N}E = 2\rho\cosh2\eta - 2 \sinh2\eta 
\int_{-\lambda_0}^{\lambda_0}[\Phi^{(1)}(\lambda,\eta) + 
\Phi^{(1)}(\lambda, 3\eta)]\rho(\lambda)d\lambda.
\eea

The other two regions are obtained from $-1 < \Delta =-\varepsilon\cosh\eta< 1$.
Denoting $\eta = i\gamma, \lambda = i\mu$ and 
instead
of (9) we have 
\bea
\lbrack\frac{\varepsilon\sinh(\mu_j + i\gamma)}{\sinh(\mu_j - i\gamma)}
\rbrack^{N-n}=
e^{-iP}\prod_{l=1}^n \frac{\sinh(\mu_j -\mu_l + i\gamma)}
{\sinh(\mu_j - \mu_l -i\gamma)}.
\eea

3) In the case $-1 < \Delta < 0, 0 < \varepsilon=+1, 0,\gamma< \pi/2$ we solved
numerically eq.(20) for $N$ up to $100$ and 
checked that state now contains only 
strings of minimal length 1, [13] i.e., all 
particle rapidities $\{\mu_j\}$ have imaginary part $i\pi/2$ (antiparticles). 
In the thermodynamic limit we then have the following integral equations for the 
distribution function
\bea
2\pi\sigma(\mu) -\int_{-\mu_0}^{\mu_0}
\Phi^{(2)}(\mu-\mu', 2\gamma)\sigma(\mu')d\mu'=\nonumber\\
-(1-\frac{n}{N})\Phi^{(3)}(\mu,2\gamma),
\eea
\bea
\int_{-\mu_0}^{\mu_0}\sigma(\mu')d\mu' =\rho,
\eea
where
\bea
\Phi^{(2)}(\mu,\gamma) = \frac{\sin\gamma}{\cosh\mu -\cos\gamma},\hspace{1cm}
\Phi^{(3)}(\mu,\gamma) = \frac{-\sin\gamma}{\cosh\mu +\cos\gamma},
\eea
and the ground-state energy is given by
\bea
\frac{1}{N}E = -2\rho\cos2\gamma + 2\sin2\gamma\int_{-\mu_0}^{\mu_0}
\Phi^{(3)}(\mu,2\gamma)\sigma(\mu)d\mu.
\eea

4) In the case $0<\Delta< 1,\varepsilon = -1, 0< \gamma <\pi/2$ 
our numerical results of (20) for lattice sizes $N$ up to 100
indicate that the ground state contains only bound pairs, 
characterized by a pair of complex particle rapidities
[14,15], like in region (15)
\bea
\mu_{\alpha}^{\pm} = \frac{1}{2}(\mu_{\alpha}^{2}\pm i\gamma)
\eea
For the density function $\sigma^{(2)}(\mu)$ we have the following integral 
equations
\bea
2\pi\sigma^{(2)}(\mu) +\int_{-\mu_0^{(2)}}^{\mu_0^{(2)}}
[2\Phi^{(2)}(\mu-\mu';2\gamma) + 
\Phi^{(2)}(\mu - \mu';4\gamma)]\sigma^{(2)}(\mu')d\mu'= \nonumber\\
(1-\rho)[\Phi^{(2)}(\mu,\gamma) + \Phi^{(2)}(\mu,3\gamma)] 
\eea
\bea
\int_{-\mu_0^{(2)}}^{\mu_0^{(2)}}\sigma^{(2)}(\mu')d\mu' = \frac{1}{2}\rho,
\eea
and the ground-state energy is given by
\bea
\frac{1}{N}E = 2\rho\cos2\gamma - 2\sin2\gamma\int_{-\mu_0^{(2)}}^{\mu_0^{(2)}}
[\Phi^{(2)}(\mu,\gamma)+\Phi^{(2)}(\mu,3\gamma)]
\sigma^{(2)}(\mu)d\mu.
\eea

Solving numerically the corresponding integral equations in regions 2, 3 and 4
we show in Fig.1 the ground-state energy as a function of 
density for some values of $\Delta$. Except region I in all regions we expect 
gapless excitation for $\rho < 1/2$ .
In order to understand the physical properties of the model 
under consideration we shall investigate the long-distance
behavior of the correlation functions.
For this purpose we shall use two-dimensional conformal field theory
[16,17] and analytic methods [18] to extract finite-size corrections from the 
Bethe-ansatz equations. The results of these calculations 
indicate that the critical fluctuations are described by a conformal field theory
with the central charge $c=1$.
The long-distance power-law behavior of the density-density correlation 
functions is given by the general form
\bea
\left<\rho(r)\rho(0)\right>\simeq\rho^2+A_1r^{-2}+A_2r^{-\alpha}
\cos(2k_Fr); \hspace{1cm}     2k_F=\pi\rho;
\eea
\bea
\rho(r)=\sigma_j^+\sigma_{j}^-, \nonumber\\
\eea
while the pair correlation function is given by
\bea
G_{\rho}(r)=\left<\sigma_{j}^+\sigma_{j+1}^+,
\sigma_{j+r}^-\sigma_{j+r+1}^-
 \right>\simeq Br^{-\beta} .
\eea
The exponents $\alpha$ and $\beta$ describing 
the algebraic decay are calculated from the dressed charge function
$\xi_0 = \xi(\lambda_0) $ 
\bea
\beta = \alpha^{-1} = \frac{1}{2[\xi(\lambda_0)]^2}
\eea
This function is obtained by the solution of the integral equations (16), 
(21) and (26) with the right-hand side replaced by $(1-\rho)$.
In Fig.2 we show the exponent $\beta$ as a function of $\rho$ for
several values of $\Delta$ in regions 2 and 4. Our numerical
results indicate that as $\rho \rightarrow 1/2$ the exponent $\beta$
tends toward the value $4\gamma/\pi$ and $\beta = 8(\pi -2\gamma)/\pi$
in regions (2) and (4), respectively.
In the region with dominant density-density correlations $\beta>1$ 
the particles prefer to move individually, instead by pair hopping,
but in the region with dominant pair correlations, $\beta<1$ they 
create two-atom molecules. For arbitry values of $\Delta > 1$ there 
exists a curve $\Delta = \Delta_0(\rho)$ separating both behaviors. 
An analogous behavior of correlation functions can be observed 
in the models [7,8] which is translated in a strong tendency to the
superconductivity.
In region 3 we have no pairs . In Fig.3 we show the exponent $\beta$
which describes now the spin-spin correlation function 
$<\sigma_j^+\sigma_{j+r}^->$. Our numerical results indicate that as 
$\rho \to 1/2$ the exponent $\beta$ tends toward the value
$\beta =4\gamma/\pi$.

We thank A. Lima Santos for useful conversations and one of us (R.Z.B.) thanks 
Laboratoire de Physique Th\'eorique des Liquides, Universit\'e Pierre et Marie 
Curie, where part of this work was done, for their hospitality.
This work was supported in part by Conselho Nacional de Desenvolvimento 
Cient\'{\i}fico e Tecnol\'ogico  - CNPq - Brazil, 
and by FINEP -  Brazil.

\newpage

{\bf REFERENCES}

\begin{enumerate}

\item F. H. L. Essler, V. E. Korepin, Exactly solvable models of strongly 
correlated electrons, World Scientific, 1994.
\item M. Gaudin, La Fonction d'Onre Bethe (Masson, Paris, 1983).
\item  P. Schlottmann, Phys.Rev. B {\bf 36}, 5177 (1987).
\item P. A. Bares and G. Blatter, Phys.Rev.Lett. {\bf 64}, 2567 (1990).
\item  E. H. Lieb and F. Y. Wu, Phys.Rev.Lett. {\bf 20}, 1445 (1968).
\item R. Z. Bariev, J. Phys. A {\bf 24},  L549 (1991).
\item R. Z. Bariev, A. Kl\"umper, A. Schadschneider and J. Zittartz,
 J.Phys.A {\bf 26}, 1249 4863 (1993) .
\item  R. Z. Bariev, A. Kl\"umper, A. Schadschneider and J. Zittartz,
  Phys.Rev. B {\bf 50},  9676 (1994); J. Phys. A {\bf 28}, 2437 (1993);
  Europhys.Lett. {\bf 32}, 85 (1995).
\item R. J. Baxter, Exactly solved models in statistical mechanics
    (Academic Press, New York, 1982).
\item A. B. Zamolodchikov and V. Fateev, Sov. J. Nucl. Phys. {\bf 32}, 298
(1980).
\item A. G. Izergin and V. E. Korepin, Commun.Math.Phys. {\bf 79}, 303 (1981).
\item G. Albertini, V. E. Korepin and A. Schadschneider, J.Phys.A {\bf 28}, L303
(1995).
\item F. C. Alcaraz and M. J. Martins, Phys.Rev.Lett. {\bf 63}, 708 (1989).
\item H. M. Babujian, Nucl. Phys. B{\bf 215 [FS7]}, 317 (1983).
\item A. N. Kirillov and N. Yu. Reshetikhin, J. Phys. A {\bf 20}, 1565 (1987).
\item J. L. Cardy, Nucl. Phys. B {\bf 270 [FS16]}, 186 (1986).
\item N. M. Bogoliubov and  V. E. Korepin, Int. J. Mod. Phys. B {\bf 3}, 
427 (1989).
\item F. Woynarovich, H. P. Eckle and T. T. Truong, 
J. Phys. A {\bf 22}, 4027 (1989).  
\end{enumerate} 
\newpage

\begin{figure}[tbp]
\caption{The ground-state energy as a function of 
the density $\rho$ for some values of 
$\Delta = -\varepsilon\cosh\eta$, in region 2,3 and 4. 
a) $\Delta = 11.5920$, b)$\Delta = 2,5092$, c)$ \Delta = 0.7071$,
d) $\Delta = 0.1423$, e) $\Delta = -0.9239$, f) $\Delta = -0,8090$,
g)$ \Delta = -0.5$}
\end{figure}
\begin{figure}[tbp]
\caption{The exponent $\beta$ describing the pair -pair correlation
function as a function of 
the density $\rho$ for some values of 
$\Delta= -\varepsilon\cosh\eta$ , in regions 2 and 4.
a)$ \Delta = 11.5919$, b)$ \Delta = 1.2039$, c) $\Delta = 0.7071$,
$\Delta = 0.3827$, e) $\Delta = 0.1423$, f)$ \Delta = 0.0383$}
\end{figure}
\begin{figure}[tbp]
\caption{The exponent $\beta$ describing the spin-spin correlation
function as a function of 
the density $\rho$ for some values of 
$\Delta= -\varepsilon\cosh\eta$ , in region 3.
a) $\Delta = -0.5$, b) $\Delta = -0.7071$, c) $\Delta = -0.8090$,
$\Delta = -0.9239$, e)$ \Delta = -0.9511$, f)$ \Delta = 0.9980$}
\end{figure}

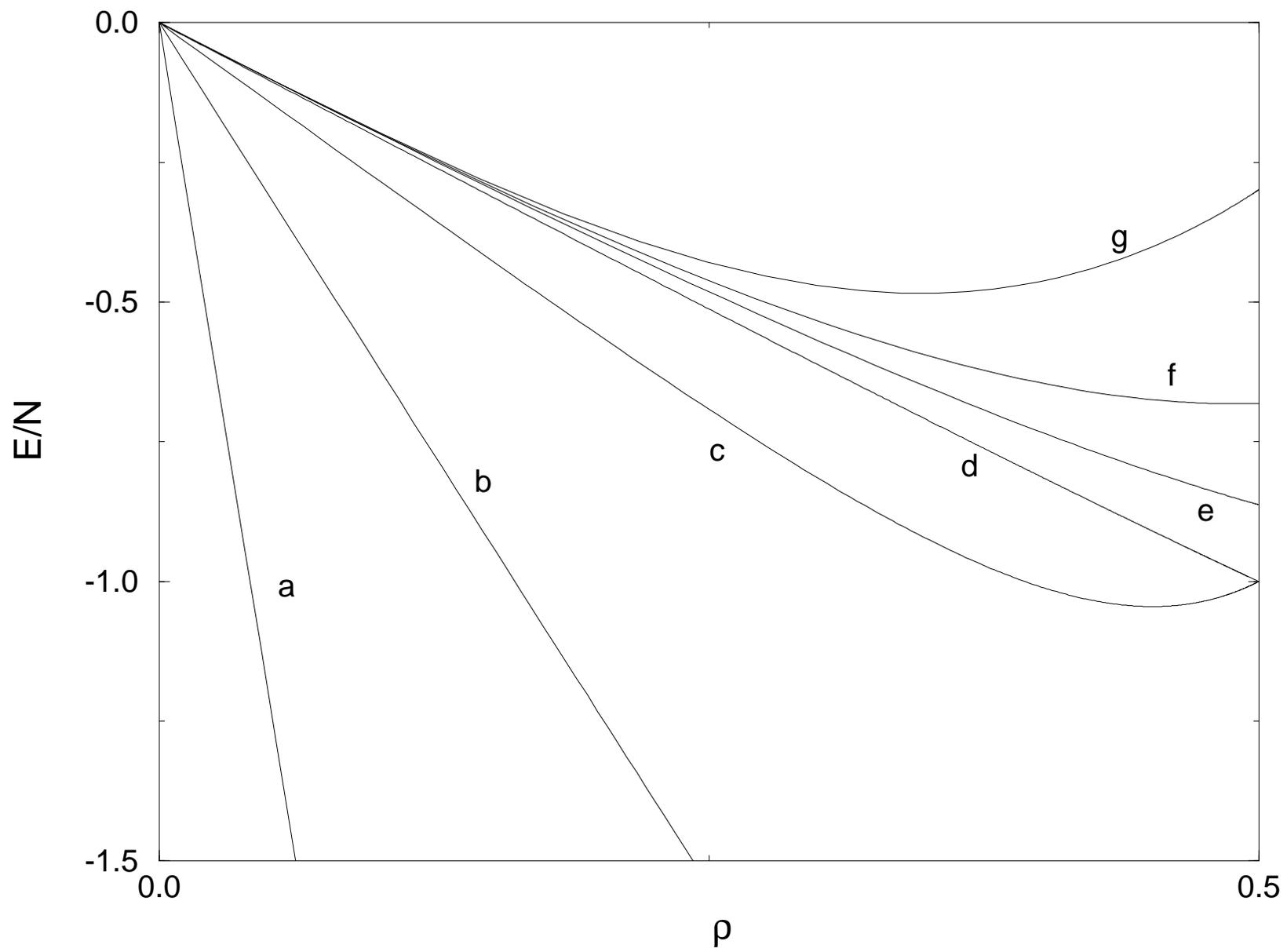
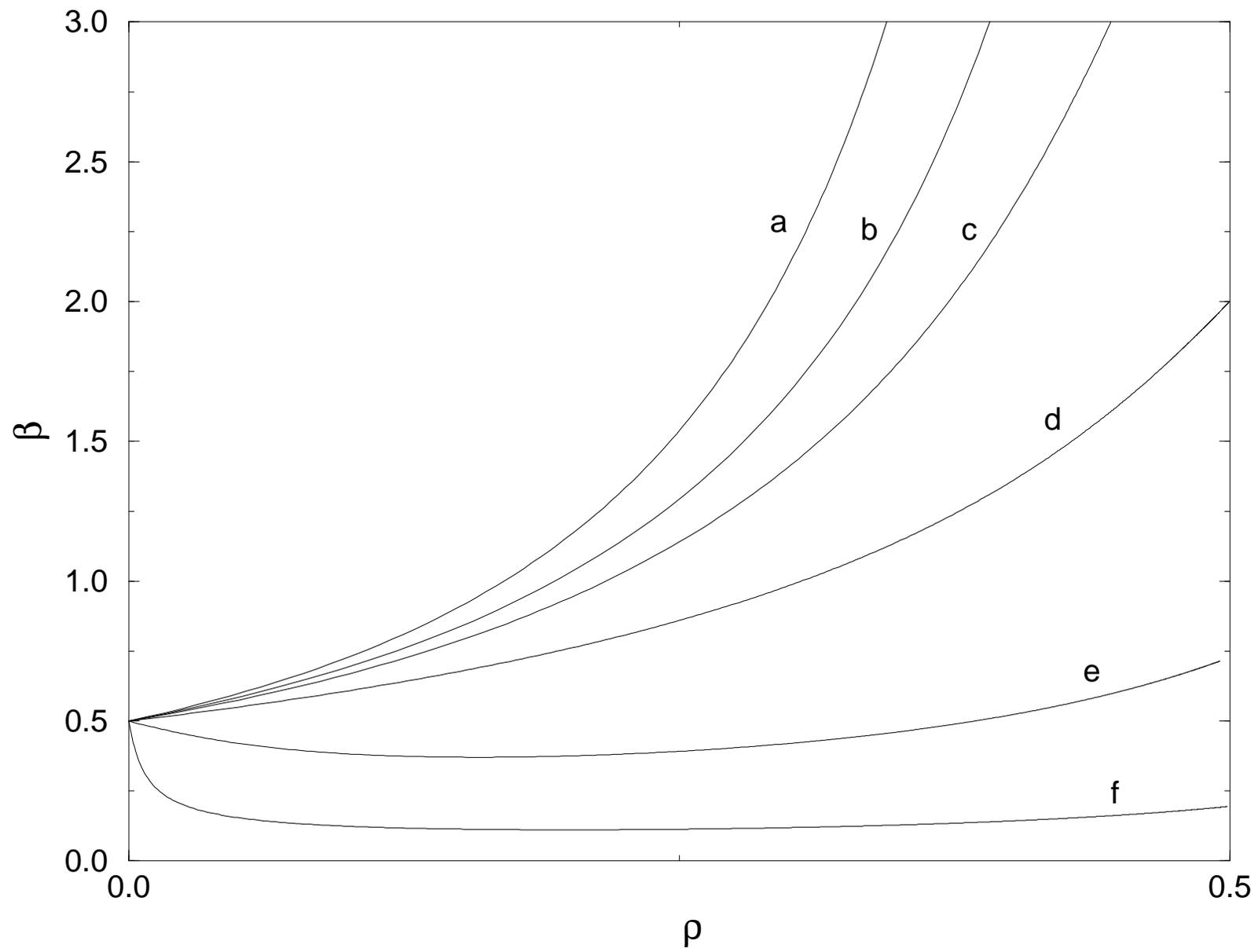
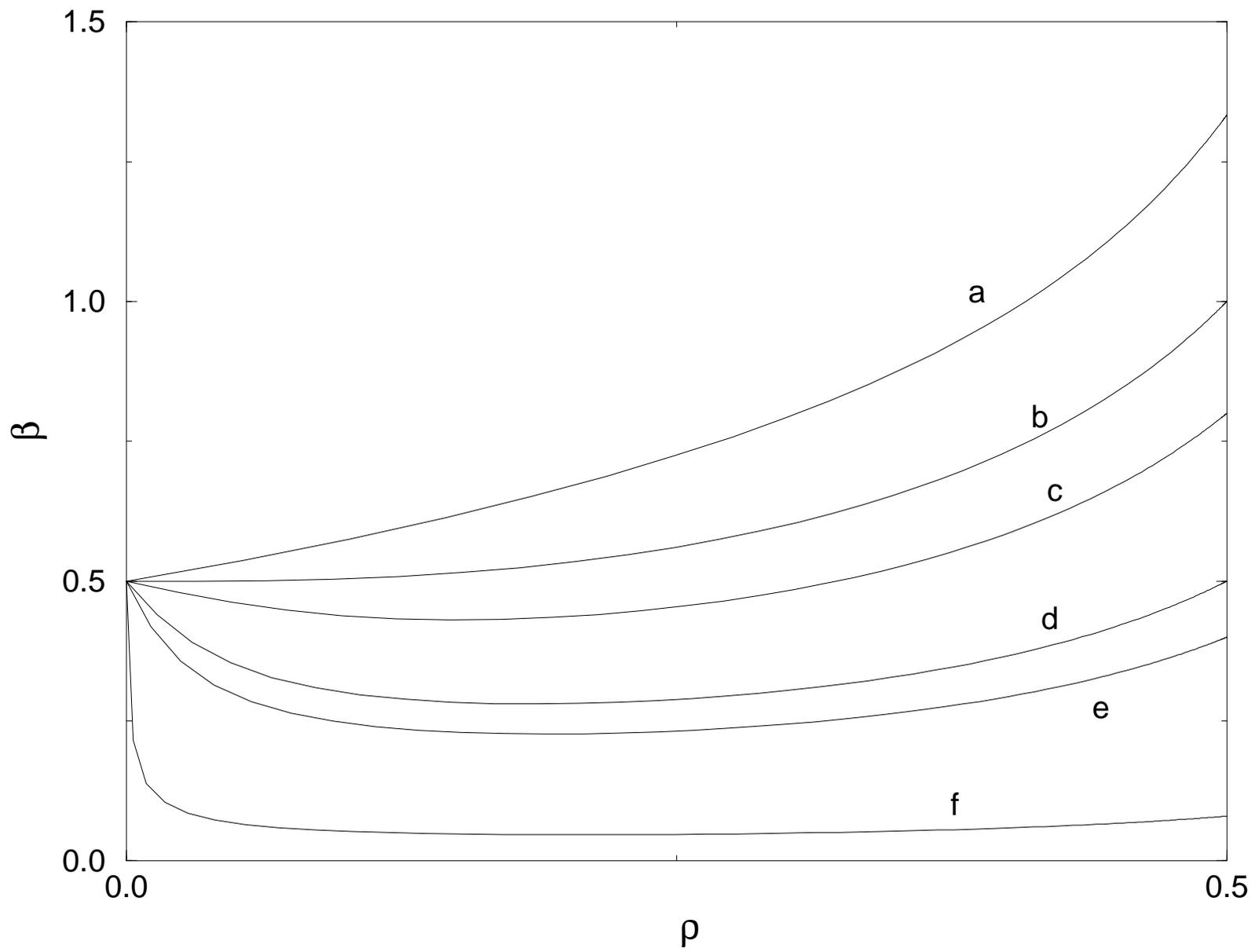
\end{document}